\documentclass[pss]{wiley2sp} 
\usepackage{amsmath}

\begin{document} 

\title{Influence of quantum critical point of first order valence transition 
on Ce- and Yb-based heavy fermions}

\titlerunning{Influence of quantum critical point of valence transition}

\author{%
  Shinji Watanabe\textsuperscript{\Ast,\textsf{\bfseries 1}},
  Kazumasa Miyake\textsuperscript{\textsf{\bfseries 2}}}

\authorrunning{Shinji Watanabe et al.}

\mail{$^{**}$Present address: Division of Materials Physics, Department of Materials Engineering Science, Graduate School of
Engineering Science, Osaka University, Toyonaka, Osaka 560-8531, Japan}

\institute{%
  \textsuperscript{1}\,Department of Applied Physics, University of Tokyo, Hongo 7-3-1, 
Bunkyo-ku, Tokyo, 113-8656, Japan$^{**}$ \\
  \textsuperscript{2}\,Division of Materials Physics, Department of Materials Engineering Science, Graduate School of
Engineering Science, Osaka University, Toyonaka, Osaka 560-8531, Japan}

\received{10 June 2009, revised 14 January 2010, accepted 14 January 2010} 
\published{online 17 February 2010} 

\pacs{71.10.-w, 71.20.Eh, 71.27.+a, 74.70.Tx, 75.40.-s} 

\abstract{%
%
%
%
\abstcol{%
Influence of quantum critical point (QCP) of first-order valence transition (FOVT) on Ce- and Yb-based heavy fermions is discussed as a key origin of anomalies such as non-Fermi liquid, metamagnetism, and unconventional superconductivity. Even in intermediate-valence materials, the QCP of the FOVT is  shown to be induced by applying the magnetic field, which creates a new characteristic energy distinct from the Kondo temperature.
  }{%
It is stressed that the key concept is closeness to the QCP of the FOVT by pointing out that the proximity of the QCP explains sharp contrast between X=Ag and X=Cd in YbXCu$_4$, field-induced valence crossover in X=Au, and field-induced first order transition as well as non-Fermi-liquid critical behaviours in CeIrIn$_5$. The valence fluctuations can be a key origin of unresolved phenomena in this family of materials.
}
}
%
%

\maketitle   

\section{Introduction}
Quantum critical phenomena in itinerant fermion systems have been intensively studied in the context of spin fluctuations~[1-3]. Recently, quantum critical phenomena in charge degrees of freedom have attracted much attention~[4]. One of such highlights is valence transition, which is a phase transition with a valence of materials element showing a discontinuous jump. A typical example is known as the $\gamma$-$\alpha$ transition in Ce metal, where the isostructural first-order valence transition (FOVT) occurs in temperature-pressure $(T,P)$ phase diagram~[5] (see Fig.~1(a)). The valence of Ce changes discontinuously between Ce$^{+3.03}$ ($\gamma$  phase) and Ce$^{+3.14}$ ($\alpha$ phase) at $T=300$ K~[6]. The FOVT line terminates at the critical end point at $(T, P)=(600~{\rm K}, 2~{\rm GPa})$. As diverging density fluctuations at the critical end point in the liquid-gas transition, valence fluctuations diverge at the critical end point. In Ce metal, the critical-end temperature is so high that quantum criticality has not been discovered. 

\begin{figure}
\includegraphics[width=70mm]{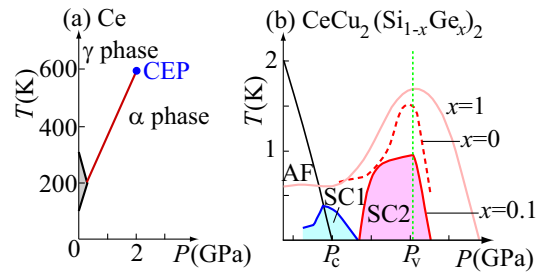}
\caption{(color online) 
 $T$-$P$ phase diagram of (a) Ce and (b) CeCu$_2$(Si$_{1-x}$Ge$_{x}$)$_2$. (a) FOVT between the $\gamma$ and $\alpha$ phases terminates at the critical end point (CEP). (b) Unified phase diagram of $x=0$~[8], $x=0.1$~[10], and $x=1$~[7]. As for $P$ axis, $P_{\rm v}=4.5$~GPa $(x=0)$, $P_{\rm v}=5.5$~GPa $(x=0.1)$, and $P_{\rm v}=15.5$~GPa $(x=1)$ are set as the same position (see ref.~[10] for details). The dotted line at $P_{\rm v}$ is a guide for the eyes.
}
\end{figure}

When the critical-end temperature is suppressed to $T=0$~K by changing materials parameters, the quantum critical end point emerges. Since the quantum critical point (QCP) is defined as the point at which the continuous-transition temperature touches $T=0$~K, we refer to the quantum critical end point as the QCP in this paper. 
The quantum criticality and electron instabilities arisen from the QCP of the FOVT has been recognized as emergence of two superconducting domes in the $T$-$P$ phase diagram of CeCu$_2$Ge$_2$~[7], CeCu$_2$Si$_2$~[8,9], and CeCu$_2$(Si$_x$Ge$_{1-x}$)$_{2}$~[10,11]: Near the QCP $P=P_{\rm c}$ where the Neel temperature is suppressed under pressure, a superconducting phase emerges (SC1). Interestingly, further enhancement of the superconducting-transition temperature $T_{\rm SC}$ emerges by further applying pressure (SC2) (see Fig.~1(b)). Around the pressure $P=P_{\rm v}$ where $T_{\rm SC}$ has a maximum in the SC2, the quantum critical behaviours such as a $T$-linear resistivity and a remarkable enhancement of residual resistivity have been observed. Since the resistivity shows the $T^{1.5}$ dependence near $P=P_{\rm c}$, i.e., the criticality predicted theoretically for the antiferromagnetic (AF) QCP in three spatial dimension~[1-3], these anomalies near $P=P_{\rm v}$ should be ascribed to a distinct origin of AF quantum criticality.
 
    Theoretically, a possibility of emergence of valence criticality at $P=P_{\rm v}$ has been pointed out in ref.~[12]: A key experimental feature is the pressure dependence of the $T^2$ coefficient $A$ estimated from the low-temperature part of resistivity~[4]. A remarkable fact is that $A$ sharply decreases with 2-3 orders of magnitude, when $P$ increases and across $P=P_{\rm v}$~[7-11]. Since $A$ is proportional to $\gamma_{\rm e}^2$ with $\gamma_{\rm e}$ being Sommerfeld constant by so-called Kadowaki-Woods relation~[13], this suggests that the effective mass of electrons $m^{*}/m_{0}$ is drastically reduced at $P=P_{\rm v}$. This sharp crossover is also reflected in the sharp change of the Kadowaki-Woods ratio between strong-correlation regime $A/\gamma_{\rm e}^2=10^{-5}$ $(P<P_{\rm v})$ and weak-correlation regime $A/\gamma_{\rm e}^2=10^{-6}$ $(P>P_{\rm v})$. If we remind the fact that the effective mass is expressed by the $f$-electron number per site in Ce (and also Yb for the $f$-hole number) based heavy fermion systems: $m^{*}/m_{0}\propto(1-n_{f}/2)/(1-n_{f})$ shown by the Gutzwiller arguments for periodic Anderson model~[14], the above observations imply that the $f$-electron number is sharply deviated from $n_{f}\sim 1$ when $P$ exceeds $P_{\rm v}$. In other words, the valence of Ce increases sharply from nearly trivalent 3+ $(n_{f}\sim 1)$ realized in the low-pressure regime to a larger valence $3+\delta$ $(n_{f}\sim 1-\delta)$ with $\delta$ being a positive value. 
    
    Recent $^{63}$Cu NQR measurement in CeCu$_2$Si$_2$ under pressure has detected that near $P=P_{\rm v}\sim4.5$~GPa the NQR frequency $\nu_{\rm Q}$ starts to deviate from linear increase expected from the monotonic volume shrinkage under hydrostatic pressure~[9]. The downward deviation from the monotonic increase in the $\nu_{\rm Q}$-$P$ plane indicates that the charge distribution around the Cu site changes around $P=P_{\rm v}$, strongly suggesting that the Ce valence changes at $P=P_{\rm v}$.
    Enhanced valence fluctuations have been shown theoretically to make the impurity scattering strong, giving rise to the enhancement of the residual resitivity~[15]. The local nature of valence fluctuations has been also shown to cause the non-Fermi-liquid critical behaviour in the resistivity: The $T$-linear dependence emerges near the QCP of the FOVT~[8].  
    
In Fig.~1(b), $T_{\rm SC}$ in the SC2 starts to increase at the lower pressure than $P_{\rm v}$, i.e., just before the Ce valence increases. After the sharp valence crossover for $P>P_{\rm v}$, the superconductivity is immediately suppressed. This feature is reproduced by the slave-boson mean-field theory taking into account of Gaussian fluctuations applied to the periodic Anderson model with an inter-orbital Coulomb repulsion (see Eq.~(1) below)~[16]: In the lower pressure region than $P_{\rm v}$, $T_{\rm SC}$ is enhanced. The density matrix renormalization group (DMRG) calculation in the same model in one spatial dimension also shows that the enhancement of the superconducting correlation occurs for $P<P_{\rm v}$ near the QCP~[17]. It is noted that the valence-fluctuation mediated pairing mechanism proposed in ref.~[12] obtained the firm grounds by unbiased numerical calculation~[17].

 Yb systems provide a hole analog of Ce systems, where in the $4f$ shell Yb$^{3+}$ gives 13 electrons and Yb$^{2+}$ gives fully occupied 14 electrons. Thus, in the hole picture $n_{f}=1$ $(n_{f}=0)$ is realized for Yb$^{3+}$ (Yb$^{2+}$). In Yb systems, the isostructural FOVT has been also known to occur in YbInCu$_4$~[18,19]: As $T$ decreases, the first-order transition from Yb$^{+2.97}$ $(n_{f}=0.97)$ to Yb$^{+2.84}$ $(n_{f}=0.84)$ takes place at $T=42$~K~[20].  In the series of YbXCu$_4$, when X=In is replaced to the other elements such as X=Ag~[19,21] and Au~[22], the first-order transition has not been observed. However, anomalous behaviours such as metamagnetism~[19] and emergence of new characteristic energy scale distinct from the Kondo temperature~[19,22] have been observed, which seem to be related to the enhanced valence fluctuations. 
 
The critical behaviours and electronic instabilities observed in the Ce- and Yb-based systems seem to indicate a strong influence of the QCP of the valence transition. As seen in Fig.~1(a), the FOVT terminates at the critical end point in the $T$-$P$ phase diagram in Ce metal. When the critical-end temperature is suppressed by controlling materials parameters, and enters into the Fermi degeneracy regime, the diverging valence fluctuations are considered to be coupled to the Fermi-surface instability. This combined effect of critical fluctuations and quantum fluctuations seems to be a key mechanism in understanding the anomalous Ce- and Yb-based heavy fermions. In the following sections, we show how the QCP of valence transition is controlled by the magnetic field, and discuss its influence on Ce- and Yb-based systems in the subsequent section. 

\section{Field-induced QCP of valence transition}
We analyze the magnetic field dependence of the QCP of the FOVT on the basis of the extended periodic Anderson model as a simplest minimal model for Ce- and Yb-based heavy fermions:
\begin{eqnarray}
  H = H_c+ H_f+ H_{\rm hyb}+ H_{U_{fc}}-h\sum_{i}(S_{i}^{fz}+ S_{i}^{cz})
\end{eqnarray}
where $H_{c}=\sum_{{\bf k}\sigma}c_{{\bf k}\sigma}^{\dagger}c_{{\bf k}\sigma}$, 
$H_{f}=\sum_{i\sigma }f_{i\sigma}^{\dagger}f_{i\sigma}+U\sum_{i}f_{i\uparrow}^{\dagger}f_{i\uparrow}f_{i\downarrow}^{\dagger}f_{i\downarrow}$, $H_{\rm hyb}=\sum_{i\sigma}(f_{i\sigma}^{\dagger}c_{i\sigma}+c_{i\sigma}^{\dagger}f_{i\sigma})$, and 
$H_{U_{fc}}=U_{fc}\sum_{i\sigma\sigma'}f_{i\sigma}^{\dagger}f_{i\sigma}c_{i\sigma'}^{\dagger}c_{i\sigma'}$. The $U_{fc}$ term is the Coulomb repulsion between $f$ and conduction electrons, and is considered to play an important role in the valence transition. For example, in the case of Ce metal which exhibits the $\gamma$-$\alpha$ transition, the $4f$- and $5d$-electron bands are located at the Fermi level~[23]. Since both the orbitals are located on the same Ce site, this term cannot be neglected. In the case of YbInCu$_4$, $H_{U_{fc}}$ also plays a crucial role for the FOVT in the hole picture of Eq.~(1)~[24]. Most of Ce- and Yb-based compounds seem to have moderate values of $U_{fc}$. However, they seem to be affected by close proximity to the QCP of the FOVT as discussed below, which requires a moderate magnitude of $U_{fc}$ in Eq.~(1).

By applying the slave-boson mean-field theory~[16] to Eq.~(1), we have determined the locus of QCP of the FOVT under the magnetic field~[25]. Figure~2 shows the ground-state phase diagram in the $\varepsilon_{f}$-$U_{fc}$ plane for $D=1$, $V=0.5$, and $U=\infty$   at $n=(n_{f}+n_{c})/2=7/8$, where the conduction band is set as $\varepsilon_{\bf k}=k^2/(2m)-D$ in three spatial dimension, and $n_{c}$ is a conduction-electron number per site. At $h=0$, when $\varepsilon_{f}$ is deep enough, the Kondo state with $n_{f}=1$ is realized. When $\varepsilon_{f}$ approaches the Fermi level, $f$ electrons are moved into the conduction band via hybridization, giving rise to the mixed-valence state with $n_{f}<1$. The FOVT between them is caused by $U_{fc}$, since large $U_{fc}$ forces electrons to pour from the $f$ level into the conduction band.  The FOVT line (solid line with open triangles) terminates at the QCP (filled circle), where the valence susceptibility $\chi_{\rm v}\equiv -\partial n_{f}/\partial \varepsilon_{f}$, i.e., valence fluctuation, diverges. The dashed line represents the points where $\chi_{\rm v}$ has a maximum as a function of $\varepsilon_{\rm f}$ for each $U_{fc}$, implying that even in the valence-crossover regime valence fluctuations are well developed. 
Note that the hybridization between f and conduction electrons is always finite in the $\varepsilon_{\rm f}$-$U_{\rm fc}$ plane shown in Fig.~2. 
Hence, the Fermi surface is always large~\cite{17}. 

\begin{figure}
\includegraphics[width=70mm]{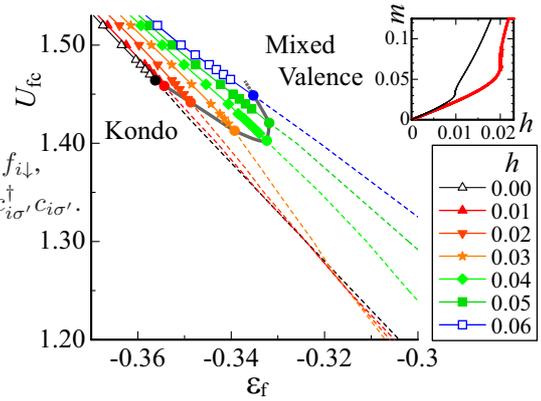}
\caption{(color online) 
 Ground-state phase diagram in the $\varepsilon_{f}$-$U_{fc}$ plane for $D=1$, $V=0.5$, and $U=\infty$ at $n=7/8$. The FOVT line with a QCP for $h=0.00$ (open triangle), $h=0.01$ (filled triangle) $h=0.02$ (filled inverse triangle), $h=0.03$ (filled star), $h=0.04$ (filled diamond), $h=0.05$ (filled square) and $h=0.06$ (open square). The gray line connects the QCP's under $h$, which is a guide for the eyes. The dashed lines represent the valence-crossover points at which $\chi_{\rm v}$ has a maximum as a function of $\varepsilon_{f}$ for each $U_{fc}$. Inset: $m$-$h$ curve for $(\varepsilon_{f},U_{fc})=(-0.354,1.458)$ (thin line) and (-0.349,1.442) (bold line). $m$ is defined by 
 $m\equiv\sum_{i}(\langle S_i^{fz}\rangle+\langle S_i^{cz}\rangle)/N$. 
}
\end{figure}

Under the magnetic field, the QCP shows non-monotonic field dependence as shown in Fig.~2: The QCP shifts to the smaller-$U_{fc}$ and larger-$\varepsilon_{f}$ direction for small $h$, and upturn behaviour emerges for $h\ge 0.04$. Since the Kondo temperature at the QCP $(\varepsilon_{f}^{\rm QCP},U_{fc}^{\rm QCP})=(0.356, 1.464)$  under $h=0$ is estimated as $T_{\rm K}^{\rm QCP}=0.074$, the characteristic field of the beginning of the upturn roughly corresponds to $T_{\rm K}^{\rm QCP}$. The origin of the upturn is due to the broadening of the spectral weight of $f$- and conduction-electrons Green function by the magnetic field larger than $T_{\rm K}^{\rm QCP}$~[25].  If Ce- and Yb-based materials are located just on the locus of QCP's for $h>0$ (a gray line) in Fig.~2, the metamagnetism will be observed at $h=h_{\rm m}$ (as seen in inset of Fig.~2). It is remarked that at the field-induced valence QCP, not only valence susceptibility $\chi_{\rm v}$, but also the magnetic susceptibility $\chi_{\rm s}\equiv -\partial m/\partial h$ diverges. Thus, uniform spin fluctuations at $k=0$ are also enhanced. 

It is noted that the field-induced QCP of the FOVT has also been confirmed by the DMRG calculation in Eq.~(1) in one dimension~[25]. This indicates that not only unbiased calculation reproduces the above result, but also a difference of spatial dimensions does not alter the main conclusion. This is ascribed to the locality of the valence transition, which has a local atomic origin~[26]. 

The metamagnetic field $h=h_{\rm m}$ corresponds to the difference of $T_{\rm K}$ at the QCP between $h=0$ and $h=h_{\rm m}$: $h_{\rm m}\sim T_{\rm K}^{\rm QCP}(h\ne 0)-T_{\rm K}^{\rm QCP}(h=0)$. This indicates that there appears a distinct energy scale from the Kondo temperature, which is characterized by the closeness to the QCP of the FOVT. As mentioned above, Ce metal is considered to have a considerable value of $U_{fc}$ due to its onsite origin. Hence, it seems to be located at $U_{fc}>U_{fc}^{\rm QCP}$ in Fig.~2, giving rise to the FOVT. On the other hand, in Ce-based compounds, the energy bands located at the Fermi level usually consist of $4f$ orbitals at the Ce site and non-$4f$ orbitals at the non-Ce site. For example, $4f$ orbitals at the Ce site and $3p$ $(4p)$ orbitals at the X site are responsible for the bands in CeCu$_2$X$_2$ (X=Si (Ge)). Then, $U_{fc}$ has the intersite origin in Ce compounds. Hence, most of Ce- and also Yb-based compounds are considered to be located in the intermediate valence-crossover regime for $U_{fc}<U_{fc}^{\rm QCP}$ in Fig.~2. This is consistent with the experimental fact that most of those compounds do not show the FOVT, but merely show the valence crossover. However, the results revealed in Fig.~2 indicate that even such compounds can be affected by the proximity to the valence QCP by applying the magnetic field. In the next section, we discuss how this newly clarified mechanism resolves the outstanding puzzles measured in Ce- and Yb-based compounds so far. 

\section{Comparison with experiments}

\subsection{YbAgCu$_4$ and YbCdCu$_4$}
YbAgCu$_4$ and YbCdCu$_4$ have paramagnetic-metal ground states. The analysis of $T$ dependence of uniform magnetic susceptibilities of YbAgCu$_4$ and YbCdCu$_4$ concluded that both have nearly the same Kondo temperatures, about 200~K~[19]. However, only in YbAgCu$_4$ a peak structure appears in the magnetic susceptibility around $T\sim 40$~K, but in YbCdCu$_4$ a conventional Pauli paramagnetism appears~[19]. Such a contrast has been also detected in the magnetic-field response: YbAgCu$_4$ shows a metamagnetic increase of the magnetization around $h\sim 40$~T, while YbCdCu$_4$ does not~[19]. These observations suggest that there exist a distinct energy scale from the Kondo temperature. 

Our results explain this sharp contrast. Figure~3(a) shows the schematic contour plot of the $f$-hole number per site, $n_{f}$, which can be also regarded as the contour plot of the Kondo temperature, $T_{\rm K}\propto (1-n_{\rm f})/(1-n_{\rm f}/2)$~[14]. In the small (large) $\varepsilon_{f}$ and $U_{fc}$ regime, the Kondo (mixed valence) state with $n_{f}=1$ $(n_{f}<1)$, i.e., small (large) $T_{\rm K}$ is realized. Since YbAgCu$_4$ and YbCdCu$_4$ have nearly the same $T_{\rm K}$, both are considered to be located in the same contour area (see Fig.~3). If YbAgCu$_4$ is closer to the QCP with a distance about $h\sim 40$~T than YbCdCu$_4$ which is less close to the QCP, the metamagnetic behaviour in YbAgCu$_4$ at $h\sim 40$~T can be naturally understood. 

\begin{figure}
\includegraphics[width=70mm]{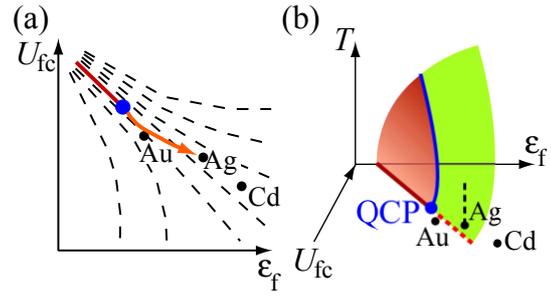}
\caption{(color online) 
(a) Schematic contour plot of $n_{\rm f}$, i.e., the Kondo temperature $T_{\rm K}$ in the $\varepsilon_{f}$-$U_{fc}$ plane for the model~(1) at $h=0$. The FOVT line (solid line) terminates at the QCP (filled circle). The QCP reaches YbAgCu$_4$ by applying $h$ about 40~T represented by an arrow. YbCdCu$_4$ is located far away from the QCP. The valence crossover surface reaches YbAuCu$_4$ by applying $h$ about 1.3~T (see text). (b) Schematic $T$-$\varepsilon_{f}$-$U_{fc}$ phase diagram at $h=0$. YbAgCu$_4$ touches the valence-crossover surface at about $T=40$~K. YbCdCu$_4$ is located too far from the QCP so that it needs too large $T$ interval to reach the valence-crossover surface.
}
\end{figure}

This picture also gives a natural explanation for the origin of the peak structure observed in the uniform susceptibility at $T\sim 40$~K in YbAgCu$_4$. Figure~3(b) illustrates a schematic phase diagram of the $T$-$\varepsilon_{f}$-$U_{fc}$ space. As shown in inset of Fig.~2, the metamagnetism with diverging magnetic susceptibility emerges at the QCP as well as the critical end points represented by the thick solid line with a circle in Fig.~3(b). An important result is that even at the valence-crossover surface extended from the FOVT surface in Fig.~3(b), the magnetic susceptibility $\chi_{\rm s}$ as well as the valence susceptibility $\chi_{\rm v}$ is enhanced~[26]. If YbAgCu$_4$ touches the valence-crossover surface at about $T_{\rm v}^{*}\sim40$~K, the observed peak in $\chi_{\rm s}(T)$ at $T\sim 40$~K is naturally explained. The reason of no conspicuous peak in $\chi_{\rm s}(T)$ in YbCdCu$_4$ can be understood as too large temperature interval to touch the valence-crossover surface (see Fig.~3(b)). Indeed, a volume expansion was observed below $T\sim 40$~K in YbAgCu$_4$~[21], which is directly related to the sharp valence crossover from Yb$^{+2.89}$ $(T>40~{\rm K})$ to Yb$^{+2.87}$ $(T<40~{\rm K})$~[27].

\subsection{YbAuCu$_4$}
Recent $^{63}$Cu NQR measurement in YbAuCu$_4$ has revealed that the characteristic temperature $T_{\rm v}^{*}$ emerges in the $T$-$h$ phase diagram as illustrated in Fig.~4(a)~[22]. Under the magnetic field for $h>1$~T, as $T$ decreases, $^{63}$Cu NQR frequency  $\nu_{\rm Q}$ decreases sharply at $T_{\rm v}^{*}(h)$. This indicates that the electric-field gradient at the Cu site changes sharply at $T_{\rm v}^{*}(h)$, implying that the charge distribution changes drastically around the Cu site. This strongly suggests that the valence of Yb changes at $T_{\rm v}^{*}(h)$. As $h$ increases and across $h_{\rm v}^{*}\sim 1.3$~T, the $T^2$ coefficient $A$ of the resistivity sharply decreases, and the residual resistivity has a cusp-like peak structure at $h_{\rm v}^{*}$~[22]. These observations also suggest that the Yb-valence crossover temperature $T_{\rm v}^{*}$ is induced by applying the magnetic field.
\begin{figure}
\includegraphics[width=70mm]{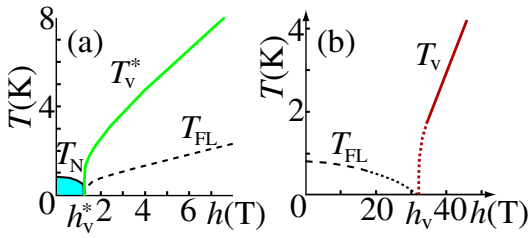}
\caption{(color online) 
(a) $T$-$h$ phase diagram of YbAuCu$_4$~[22]. The Neel temperature $T_{\rm N}$ is suppressed and the Yb valence crossover temperature $T_{\rm v}^{*}$ is induced for $h>h_{\rm v}^{*}\sim1.3$~T. (b) $T$-$h$ phase diagram of CeIrIn$_5$ $(h \parallel c)$~[33]. The first-order transition is induced for $h>h_{\rm v}$ (solid line). In (a) and (b), For $T<T_{\rm FL}$ (dashed line) the resistivity shows the Fermi liquid behaviour $\sim T^2$. In (b) dotted lines are guides for the eyes.}
\end{figure}

This can be naturally understood if YbAuCu$_4$ is located just below the locus of the QCP as illustrated in Fig.~3(a). As shown in Fig.~2, for $h<T_{\rm K}^{\rm QCP}(h=0)$, the valence-crossover line extends to the Kondo regime (dashed lines for $h=0.02$ and 0.03 cross that for $h=0$). Then, by applying $h$, the valence-crossover surface approaches and reaches the point indicated as Au in Fig.~3(a) at $h=1.3$~T, which gives $h_{\rm v}^{*}$ in Fig.~4(a). As $h$ further increases, the QCP moves as an arrow in Fig.~3(a). Hence, the valence-crossover temperature $T_{\rm v}^{*}(h)$ increases as illustrated in Fig.~4(a). In the large $h$ limit, the Yb$^{3+}$ state is expected to be realized, since the Zeeman energy gain is mostly earned by the $n_{f}=1$ state. Hence, $T_{\rm v}^{*}(h)$ is expected to start to decrease for larger $h$, and finally to touch absolute zero. Indeed, this behaviour should appear if the valence QCP's exhibit the upturn as shown in Fig.~2. 

Here, we point out the close similarity of the $T$-$h$ phase diagram between YbAuCu$_4$~[22] and YbRh$_2$Si$_2$~[28]. In YbRh$_2$Si$_2$, a very similar phase diagram as in Fig.~4(a) has been obtained with remarkable anomalies near $h\sim h_{\rm v}$ such as the $T$-linear resistivity, residual resistivity peak, changes of the Hall coefficient, magnetostriction, and magnetization~[28]. We stress here that these properties are explained if the system is close to the valence QCP, as mentioned above. Indeed, a band-structure calculation has shown that the change of the Hall coefficient around $h\sim h_{\rm v}$ in YbRh$_2$Si$_2$ can be explained by a tiny Yb valence change~[29]. Hence, it is important to examine the possibility experimentally whether the Yb valence changes at $h_{\rm v}$ in YbRh$_2$Si$_2$. 
Enhanced valence fluctuation is a possible origin of
unconventional criticality observed in YbRh2Si2~[26].

\subsection{CeIrIn$_5$}
CeIrIn$_5$ is a heavy fermion metal, which exhibits the superconducting transition at $T\sim 0.4$~K. In the $T$-$h$ phase diagram, the first-order-transition line emerges as illustrated in Fig.~4(b), which was detected by the jump in the magnetization curve~[30-33]. Capan {\it et al}. observed anomalous behaviours that the residual resistivity increases as $h$ approaches $h_{\rm v}$. Furthermore, the power of the $T$ dependence of resistivity shows a convex curve in the $T^{1.5}$ plot, which seems to indicate that the resistivity tends to show $T$-linear dependence. They pointed out that non-Fermi liquid behaviours even evident in $h=0$ may be related to the matamagnetic anomaly at $h\sim h_{\rm v}$, although its mechanism was not clarified~[33]. 

    Our results give a natural explanation for these anomalies: This is readily understood if CeIrIn$_5$ is located inside the enclosed area of the QCP line for $h\ne 0$ as in Fig.~2. Namely, at $h=0$ the system is considered to be located at the valence-crossover regime (i.e., for $U<U_{fc}^{\rm QCP}$ in Fig.~2, since no evidence of the first-order transition was observed in any physical quantities as a function of $T$ at $h=0$.) However, when $h$ is applied, the QCP of the FOVT approaches and eventually goes across, causing the metamagnetic transition in the magnetization curve. As mentioned above, near the QCP the residual resistivity is enhanced~[15] and the $T$-linear resistivity is observed in the wide-$T$ region~[8], which are quite consistent with the observations~[30-33]. Furthermore, the emergence of the first-order transition line in the $T$-$h$ phase diagram is also in agreement with the present picture. We stress that CeCoIn$_5$ which has almost the same Fermi surfaces and similar crystalline-electric-field levels as those of CeIrIn$_5$ does not show such anomalies. This indicates that the viewpoint of the closeness to the valence QCP is also indispensable in understanding the Ce115 systems. 

\section{Summary}

We clarified the mechanism of how the QCP of the FOVT is controlled by the magnetic field. Even in the intermediate valence materials, the QCP of the FOVT is shown to be induced by applying the magnetic field. Valence fluctuations developed at the valence-crossover surface near the QCP cause various anomalies. We discussed that the proximity of the valence QCP resolves the outstanding puzzles observed in YbXCu$_4$ (X=Ag, Cd, Au) and CeIrIn$_5$.  A key concept is the closeness to the QCP of the FOVT.  Influence of the valence QCP offers a key origin in understanding unresolved anomalies in Ce- and Yb-based heavy fermions.


%
%

\end{document}